\documentclass[twocolumn,10pt,amsmath,amssymb]{revtex4}
\usepackage{graphicx}

\usepackage{amsmath,amsfonts,amssymb,bm}

\begin{document}

\draft
\title{Ballistic L\'evy walk with rests: Escape from a bounded domain}

\author
{A. Kami\'nska and T. Srokowski}

\affiliation{
 Institute of Nuclear Physics, Polish Academy of Sciences, PL -- 31-342
Krak\'ow, Poland }

\date{\today}

\begin{abstract}

The L\'evy walk process for the lower interval of the time of flight distribution ($\alpha<1$) and with finite resting time 
between consecutive flights is discussed. The motion is restricted to a region bounded by two absorbing barriers and 
the escape process is analysed. By means of a Poisson equation, the total density, which includes both flying and resting phase, 
is derived and the first passage time properties determined: the mean first passage time appears proportional to the barrier position; 
moreover, the dependence of that quantity on $\alpha$ is established. Two limits emerge from the model: of short waiting time, that 
corresponds to L\'evy walks without rests, and long waiting time which exhibits properties of a L\'evy flights model. 
The similar quantities are derived for the case of a position-dependent waiting time. 
Then the mean first passage time rises with barrier position faster than for 
L\'evy flights model. The analytical results are compared with Monte Carlo trajectory simulations. 

\end{abstract} 


\maketitle

\section{Introduction}

The L\'evy walk model lets a walker move with a finite velocity, in contrast to a L\'evy flights model when 
displacements are instantaneous \cite{gei,zum,kla1,zab,froe}. 
The parameter $\alpha$ in a time of flight distribution $\tau^{-1-\alpha}$ singles out two 
qualitatively different processes: when $0<\alpha<1$ and $1<\alpha<2$. In the first case, the mean time of flight diverges resulting 
in a ballistic diffusion: the mean-squared displacement rises with time as $t^2$. Processes characterised by $\alpha$ from the 
lower interval are discussed in context of such phenomena as some properties of nanocristals \cite{brok} 
and blinking quantum dots \cite{marg}. The L\'evy walk model usually assumes that a new jump takes place immediately 
after the termination of the previous one. However, it is natural to expect that the walker may rest between consecutive jumps 
and then a finite waiting time has to be included in the model \cite{kla2,zab1,tay}. Though this version of the L\'evy walk model 
is highly realistic, it is rarely discussed. 
If walker moves in a nonhomogeneous environment the distribution of the waiting time may be position-dependent \cite{kam17,kam18}. 

The aim of this paper is to study one-dimensional ballistic L\'evy walks ($\alpha<1$), restricted to a finite interval by two absorbing barriers. 
The quantities that characterise the escape from a bounded domain are often discussed and applied in many physical problems \cite{red}. 
One asks about a time required to reach the barrier for the first time (a first passage time) and its mean $T$ (MFPT) which, if exists, 
provides a simple estimation of the escape rate. 
The properties of the escape process change after substituting instantaneous jumps by walks with 
a finite velocity which effect is especially pronounced if $\alpha<1$: 
the numerical analysis \cite{dyb17}, performed for the L\'evy walks without rests, demonstrates, in particular, that MFPT scales 
with the barrier position as $L$ while for the L\'evy flights $T\propto L^\alpha$ holds \cite{zoi,dyb17}. 
In this paper, we derive expressions for the first passage time characteristics taking into account a finite and random waiting time 
between consecutive displacements. In Section II, we define the L\'evy walk process with rests in the presence of the absorbing barriers. 
The density distribution describing that process is derived and the first passage time statistics deduced in Section III. 
The problem is generalised to the case of a position-dependent waiting time in Section IV. 

\section{Definition of the process} 

The L\'evy walk trajectory consists of a sequence of displacements when the walker moves with a constant velocity $v$. 
Before the next jump, a new direction is chosen: walker may depart to the left or to the right with the same probability. 
The time of a single flight, $\tau$, is a random variable determined by a density distribution $\psi(\tau)$ which is one-sided 
and has the asymptotics $\tau^{-1-\alpha}$, where $0<\alpha<1$. That power-law tail corresponds to the Laplace transform, 
\begin{equation}
\label{psiods}
\psi(s)=1-c_1s^{\alpha}, 
\end{equation} 
where $c_1=$ const. More precisely, we assume the following form of $\psi(\tau)$:
\begin{eqnarray} 
\label{psinu}
\psi(\tau)=\left\{\begin{array}{ll}
\alpha\epsilon^\alpha \tau^{-1-\alpha}  &\mbox{for  $\tau>\epsilon$}\\
0 &\mbox{for $\tau\le\epsilon$},
\end{array}
\right.
\end{eqnarray}  
where $\epsilon=$ const. Taking the Laplace transform from Eq.(\ref{psinu}) and comparing the result with Eq.(\ref{psiods}) 
yields $c_1$, 
\begin{equation}
\label{c1}
c_1=\lim_{s\to0}[s^{-\alpha}-\alpha\epsilon^\alpha\Gamma(-\alpha,\epsilon s)]=\epsilon^\alpha\Gamma(1-\alpha),
\end{equation}
where we applied the expansion of an incomplete Gamma function, $\Gamma(a,b)=\Gamma(a)-b^a/a+b^{a+1}/(a+1)+\dots$ \cite{ryz}. 
Since the walk-size $\xi$ is determined by $\tau$, both quantities are coupled in the jump density distribution: 
\begin{equation}
\label{jden}
\bar\psi(\xi,\tau)=\frac{1}{2}\delta(|\xi|-v\tau)\psi(\tau). 
\end{equation} 
After walker terminates its jump, and before the next direction and new time $\tau$ are sampled, it remains at rest. The resting 
time is a random quantity and follows from the exponential distribution with a rate $\nu$, then the mean waiting time is $1/\nu$. 
Both phases of the motion, namely of particles in flight and in rest, are quantified by two density distributions: 
$p_v(x,t)$ and $p_r(x,t)$, respectively. The total density, $p(x,t)=p_r(x,t)+p_v(x,t)$, 
is normalised to unity but the contribution of individual phases to the total probability may change with time: for $\alpha<1$, 
$p_r(x,t)$ decays and the flying phase prevails at long time. The time evolution of density of resting particles is governed by a 
master equation \cite{kam17}, 
\begin{equation}
\label{meq}
\begin{split}
\frac{\partial}{\partial t}&p_r(x,t)= -\nu p_r(x,t)\\ 
&+\nu\int_0^t\int p_r(x',t-t')\frac{1}{2}\psi(t')\delta(|x-x'|-vt')dt'dx'
\end{split}
\end{equation} 
and $p_v(x,t)$ is given by the integral, 
\begin{equation}
\label{pv}
p_v(x,t)=\nu\int\int_0^t\Psi(t')\delta(|x-x'|-vt')p_r(x',t-t')dx'dt',  
\end{equation} 
where $\Psi(t)=\int_t^\infty\psi(t')dt'$. 

We assume that the motion is restricted to the interval $(-L,L)$ by introducing absorbing barriers at $\pm L$ which means boundary conditions, 
\begin{equation}
\label{warbrz}
p(\pm L,t)=0.
\end{equation}
The first passage time density distribution is defined as a probability that the time needed 
to reach the barrier for the first time lies within the interval $(t,t+dt)$ \cite{red}. 
The survival probability, namely the probability that the particle never reached those barriers up to time $t$, is given by 
\begin{equation}
\label{sodt}
S(t)=\int_{-L}^L p(x,t)dx. 
\end{equation}
The first passage time density distribution reflects the change of the survival probability with time, 
\begin{equation}
\label{pfp}
p_{FP}(t)=-dS(t)/dt, 
\end{equation}
and MFPT is given by the integral, 
\begin{equation}
\label{mfpt}
T=\int_0^\infty tp_{FP}(t)dt=\int_0^\infty S(t)dt. 
\end{equation}

\section{Fractional equations and mean first passage time} 

To analyse the first passage time characteristics we need an equation for the total density $p(x,t)$ which 
satisfies the boundary conditions (\ref{warbrz}). We start from (\ref{meq}) taking the Fourier and Laplace transforms and keeping the lowest terms 
in the expansion in powers of $k$ and $s$. There are a few possibilities of passing to the limits $\{k,s\}\to \{0,0\}$ and 
the order of taking those limits may influence final density distributions and fluctuations \cite{schm}. 
In this section, we first assume a given (small) value of $k$ and next take the limit $s\to0$. 
The other order of taking the limits, namely first taking $s\to 0$, one can follow the behaviour of the density close to the origin \cite{zab}. 
However, this procedure does not lead to a diffusion equation and a mean square displacement cannot be determined. 
Then the equation for $p_r(x,t)$ reads \cite{kam18}, 
\begin{equation}
  \label{lmeq1}
sp_r(k,s)-P_0(k)=-c_1 \nu[s^{\alpha}+Bv^2k^2 s^{\alpha-2}]p_r(k,s),
  \end{equation}
where $B=\alpha(1-\alpha)/2$ and $P_0(x)$ stands for an initial condition. 
The expression determining the density of particles in flight 
follows from Eq.(\ref{pv}); the application of the Laplace transform yields, 
\begin{equation}
\label{solv10}
p_v(k,s)=c_1 \nu\left[s^{\alpha-1}-s^{\alpha-3}\frac{1}{2}(1-\alpha)(2-\alpha)v^2k^2\right]p_r(k,s). 
\end{equation}
The inversion of Eq.(\ref{solv10}) reads, 
\begin{equation}
\label{solv11} 
\begin{split} 
p_v(x,t)=&c_1 \nu\left[_0D_t^{\alpha-1}+\frac{v^2}{2}(1-\alpha)(2-\alpha) _{0}D_{t}^{\alpha-3}\frac{\partial^2}{\partial x^2}\right]\\
&\times p_{r}(x,t),
\end{split} 
\end{equation}
which is a fractional equation \cite{podl} and involves a fractional Riemann-Liouville integral defined as \cite{kilbas}, 
\begin{equation}
\label{rilo}
_0D_t^{-\beta}f(t)=\frac{1}{\Gamma(\beta)}\int_0^t dt'\frac{f(t')}{(t-t')^{1-\beta}}, 
\end{equation} 
where $\beta>0$. Note that the superscript in the above operator is negative which differentiates the above definition from 
a fractional differential operator. We apply a property that $\delta'(x)$ is an odd function to evaluate the time derivative from $p_v(x,t)$, 
using Eq.(\ref{pv}). 
\begin{equation}
\begin{split}
\label{pvdif}
&\frac{\partial p_v(x,t)}{\partial t}=\nu\int\int_0^t\Psi(t')\delta'(|x-x'|-vt')p_r(x',t-t')dx'dt'\\
&=-\nu\int\int_0^t\Psi(t')\delta(|x-x'|-vt')\partial p_r(x',t-t')/\partial t'dx'dt'.
\end{split}
\end{equation} 
Passing to the limit of small $s$ yields a Poisson equation, 
\begin{equation}
\label{solv11a} 
\begin{split} 
\frac{\partial p_v(x,t)}{\partial t}=&-c_1 \nu\left[\frac{\partial^2}{\partial t^2}+\frac{v^2}{2}(1-\alpha)(2-\alpha)\frac{\partial^2}{\partial x^2}\right]\\
&\times {_0D}_{t}^{\alpha-2}p_r(x,t).
\end{split} 
\end{equation}
for an unknown function $_{0}D_{t}^{\alpha-2}p_r(x,t)$ where lhs is regarded as a source. 
We will solve this equation with given initial and boundary conditions and then the time evolution of the total density $p(x,t)$ 
can be determined from the expression, 
\begin{equation}
\label{rozc1}
\frac{\partial p(x,t)}{\partial t}=c_1v^2\nu (1-\alpha) \frac{\partial^2}{\partial x^2}{ _0D}_t^{\alpha-2}p_{r}(x,t), 
\end{equation}
which results from the combining (\ref{lmeq1}) with (\ref{solv11a}). 


Eq.(\ref{solv11a}) will be solved by a variable separation and evaluating eigenfunctions 
corresponding to both variables. The expansion of the densities reads, 
\begin{equation}
\label{prwl}
p_r(x,t)=\sum_{n=0}^{\infty}X_n(x)T_n(t) 
\end{equation}
and 
\begin{equation}
\label{pvwl}
p_v(x,t)=\sum_{n=0}^\infty X^{(v)}_n(x)T^{(v)}_n(t).
\end{equation} 
In this way, from Eq.(\ref{solv11a}) we will obtain an equation that determine the eigenfunctions corresponding to position 
and that for the expression ${_0D}_t^{\alpha-2}p_{r}(x,t)$. Inserting (\ref{prwl}) and (\ref{pvwl}) into (\ref{solv11a}) yields for each $n$,
\begin{equation}
\label{poij}
\begin{split}
&-\frac{dT^{(v)}_n(t)}{dt}X^{(v)}_n(x)=\\
&\nu c_1\left[\frac{{\partial^2}}{\partial
t^2}+\frac{1}{2}(2-\alpha)(1-\alpha)v^2\frac{\partial^2}{\partial
x^2}\right]{_0D}_t^{\alpha-2}[X_n(x)T_n(t)].
\end{split}
\end{equation} 
The separation of variables produces an equation that determine the eigenfunctions $X_n(x)$, 
  \begin{equation}
  \label{fwx}
\frac{d^2}{dx^2}X_n(x)+\lambda_n X_n(x)=0,
   \end{equation} 
and also $X^{(v)}_n(x)$ since Eq.(\ref{solv11a}) can only be solved 
if the eigenfunctions $X_n(x)$ are of the same form as those corresponding to 
the term of nonhomogeneity. More precisely, there are two possibilities: either (a) $X^{(v)}_n(x)=-X_n(x)$ or (b) $X^{(v)}_n(x)=X_n(x)$ 
and, for the version (a), Eq.(\ref{solv11a}) yields, 
   \begin{equation}
  \begin{split}
  \label{fwt2}
&\frac{dT^{(v)}_n(t)}{dt}=\\
&\nu c_1\left[\frac{d^2}{dt^2}-\frac{1}{2}(2-\alpha)(1-\alpha)v^2{\lambda_n}\right]{_0}D_t^{\alpha-2}T_n(t).
\end{split}
   \end{equation} 
The intensities of both phases of the motion, $\phi_r(t)=\int p_r(x,t)dx$ and $\phi_v(t)=\int p_v(x,t)dx$, 
are related via Eq.(\ref{meq}) and Eq.(\ref{pvdif}); the integration over $x$ of the convolutions in those equations yields, 
\begin{equation}
\label{we4} \phi'_r(t)=\phi'_v(t)=-\nu c_1 {_0}D^{\alpha}_t\phi_r(t),
\end{equation} 
which, after inserting into Eq.(\ref{fwt2}), produces the equation, 
\begin{equation}
\label{we44} 
\sum_{n=0}^\infty \phi_n \frac{dT^{(v)}_n(t)}{dt}=-\nu c_1{_0}D_t^{\alpha}\sum_{n=0}^\infty \phi_n T_n(t),
\end{equation}
where $\phi_n=-\int X_n(x)dx$. Finally, we insert Eq.(\ref{fwt2}) into the above equation and, 
since it has to be satisfied for any choice of the basis functions $X_n(x)$, we obtain for any $n$, 
  \begin{equation}
  \label{fwt1s}
\frac{d^2}{dt^2} {_0}D_t^{\alpha-2}T_n(t)-C^2{\lambda_n}_ 0D_t^{\alpha-2}T_n(t)=0, 
   \end{equation}
where $C=v\sqrt{(1-\alpha)(2-\alpha)}/2$. 

The version (b) does not apply since it leads to unphysical results. Indeed, the counterpart of Eq.(\ref{fwt2}) reads, 
  \begin{equation}
  \begin{split}
  \label{fwt1}
&-\frac{\partial T^{(v)}_n(t)}{\partial t}=\\
&\nu c_1\left[\frac{d^2}{dt^2}-\frac{1}{2}(2-\alpha)(1-\alpha)v^2{\lambda_n}\right]{_0}D_t^{\alpha-2}T_n(t),
\end{split}
   \end{equation} 
and then ${\lambda_n}{_0}D_t^{\alpha-2}T_n(t)=0$ which, according to Eq.(\ref{rozc1}), would mean a stationary state. Therefore, we continue with the version (a). 

We solve Eq.(\ref{fwx}) with the boundary conditions (\ref{warbrz}). They imply $X_n(-L)=X_n(L)=0$ yielding the solution in the form, 
\begin{equation}
\label{fwxs}
X_n(x)=\cos(\sqrt{\lambda_n}x), 
\end{equation}
where the eigenvalues $\lambda_n=\pi^2n^2/4L^2$ ($n=1,3,5,\dots$). Inserting these eigenvalues into Eq.(\ref{fwt1s}) and solving the equation yields, 
\begin{equation}
\label{fwxt}
_0D_t^{\alpha-2}T_n(t)=a_n\hbox{e}^{-C \sqrt{\lambda_n}t}+b_n\hbox{e}^{C \sqrt{\lambda_n}t},
\end{equation}
and the solution of Eq.(\ref{poij}) is 
\begin{equation}
\label{poijs}
\begin{split} 
&_0D_t^{\alpha-2}[X_n(x)T_n(t)]=\left[a_{2n+1}\exp(-\frac{C\pi(2n+1)t}{2L})\right.\\
&\left.+b_{2n+1}\exp(\frac{C\pi(2n+1)t}{2L})\right]\cos\frac{\pi(2n+1)x}{2L}.
\end{split} 
\end{equation} 
To evaluate the total density $p(x,t)$, we sum the above result over $n$ and insert into Eq.(\ref{rozc1}), 
\begin{equation}
\label{solt0}
\begin{split} 
&p(x,t)=-\frac{c_1\nu}{2}\sqrt{\frac{1-\alpha}{2-\alpha}}\sum_{n=0}^\infty\left[a'_{2n+1}\exp(-\frac{C\pi(2n+1)t}{2L})\right.\\
&\left.+b'_{2n+1}\exp(\frac{C\pi(2n+1)t}{2L})\right]\cos\frac{\pi(2n+1)x}{2L}, 
\end{split} 
\end{equation} 
where the new coefficients $a'_{2n+1}$ and $b'_{2n+1}$ can be determined from the following conditions: 
$p(x,\infty)=0$, which implies $b'_{2n+1}=0$, and the initial condition $p(x,0)=\delta(x)$, which, after taking into account 
the orthonormality of the cosine function, implies $a'_{2n+1}=-\frac{4}{Lc_1\nu}\sqrt{\frac{2-\alpha}{1-\alpha}}$. 
The final expression for the total density reads, 
\begin{equation}
  \label{solt}
p(x,t)=\frac{1}{L}\sum_{n=0}^{\infty}\exp(-\frac{C\pi(2n+1)t}{2L})\cos\frac{\pi(2n+1)x}{2L}.
  \end{equation} 
\begin{figure}
\includegraphics[width=90mm]{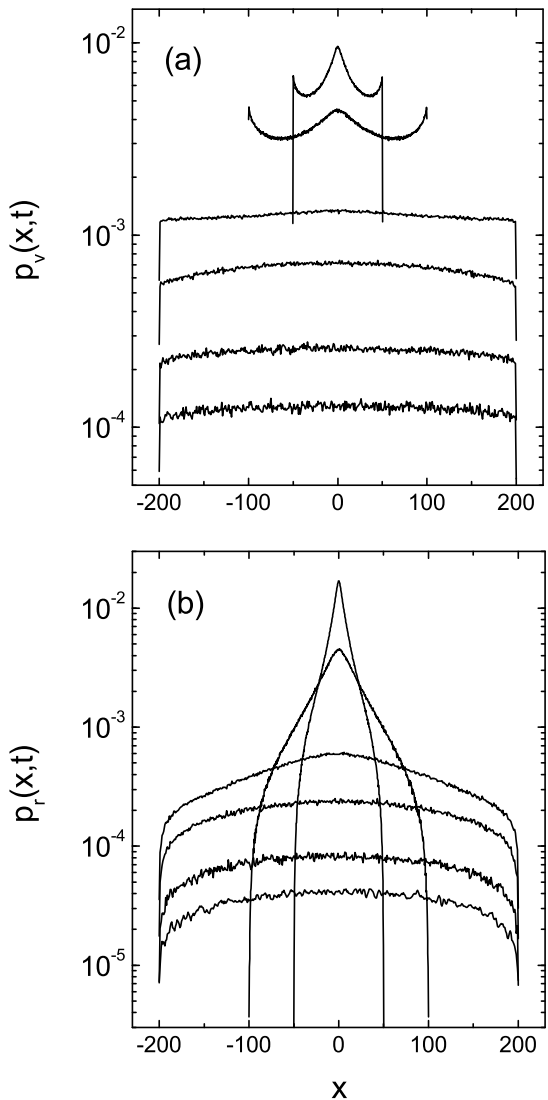}
\caption{Time evolution of density distribution for both phases of the motion, (a) $p_v(x,t)$ and (b) $p_r(x,t)$, 
calculated from trajectory simulations for $\alpha=0.5$, $L=200$ and the following values of time: 
50, 100, 300, 500, 800 and 1000 (from top to bottom in the centre of the figures). $10^7$ trajectories for each curve were calculated.}
\end{figure}  

On the other hand, the density distributions can be obtained from numerical simulation of individual trajectories. In those calculations, 
we sample the waiting time from the exponential distribution with the rate $\nu$ and the time of flight from the power-law 
distribution, according to Eq.(\ref{psinu}). Fig.1 presents a time evolution of the density distribution for both phases of the motion. 
If time is shorter than the time needed to 
reach the barrier ($t<L/v$), for $p_v(x,t)$ we observe (beside remnants of the decaying initial distribution) peaks at $x=\pm t$; 
those peaks are the solution of the wave equation (cf. Eq.(26) in \cite{kam18}). When time exceeds the value $L/v$ peaks are absorbed, 
the wave equation no longer governs the distribution and $p_v(x,t)$ becomes flat. 
\begin{figure}
\includegraphics[width=92mm]{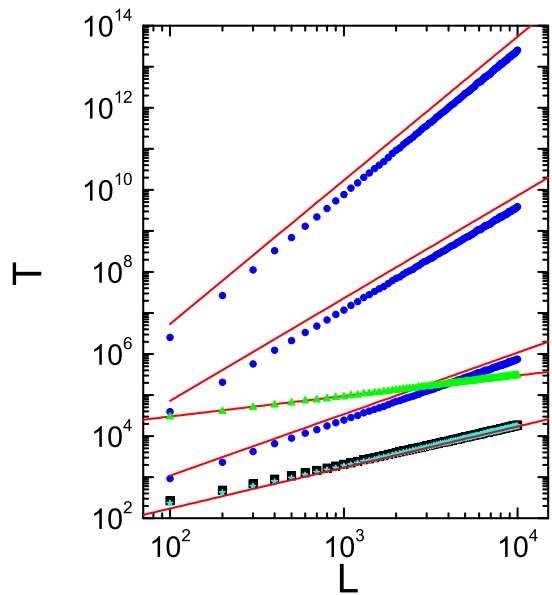}
\caption{MFPT as a function of barrier position for $\alpha=0.5$. Results presented as black squares (the lowest set of points) were calculated from trajectory simulations with $\nu=1$ and the red solid line corresponds to Eq.(\ref{mfptf}). 
Green triangles follow from calculations with $\nu=10^{-3}$ and obey a dependence $L^\alpha$, marked by the red solid line. 
Cyan stars correspond to numerical calculation for $\theta=-0.4$ and actually coincide with black squars. 
The blue points (from bottom to top) correspond to $\theta=$1, 2 and 3; the red solid lines (tracing the blue points) mark the dependence (\ref{mfptfth1}).}
\end{figure}

The survival probability follows from a direct integration of $p(x,t)$ over $x$, 
\begin{equation}
  \label{sodtf}
S(t)=\frac{4}{\pi}\sum_{n=0}^{\infty}\frac{(-1)^n}{2n+1}\exp(-\frac{C\pi (2n+1)t}{2L}), 
  \end{equation}
and we conclude from Eq.(\ref{sodtf}) that the decay pattern at large time is exponential. The differentiation of $S(t)$ 
and summation of the series yields the first passage time density, 
\begin{equation}
  \label{fptd}
p_{FP}(t)=\frac{C}{L\cosh(C\pi t/2L)}, 
  \end{equation}
while the integration of $S(t)$ yields MFPT, 
  \begin{equation}
  \label{mfptf}
T=\frac{8L}{C\pi^2}\sum_{n=0}^{\infty}\frac{(-1)^n}{(2n+1)^2}=\frac{16L}{v\pi^2}\frac{{\cal G}}{\sqrt{(1-\alpha)(2-\alpha)}}, 
  \end{equation}
where ${\cal G}=0.916\dots$ is a Catalan constant \cite{ryz}. 
The above result is compared with numerical calculations in Fig.2, where 
the dependence $T\propto L$ is illustrated, and in Fig.3 for the dependence $T(\alpha)$. For $\nu=1$ and the range of $L$ 
taken into account in Fig.3, the results of the simulations do not agree with 
Eq.(\ref{mfptf}) at large $\alpha$ while we observe a good agreement for the entire interval $\alpha\in(0,1)$ in the limit $\nu\to\infty$. 
This limit corresponds to the L\'evy walk process without rests for which the relation $T\propto L$ is well-known \cite{dyb17}. 
On the other hand, the limit $\nu\to0$ means a long waiting time compared to a mean time walker needs to arrive at the barrier 
and this case corresponds to the L\'evy flight process. Indeed, Fig.2 shows that then one observes the scaling $T\propto L^\alpha$ \cite{zoi}. 
The relation $T\propto L$ still holds for small $\nu$ but at larger values of $L$. 
\begin{figure}
\includegraphics[width=95mm]{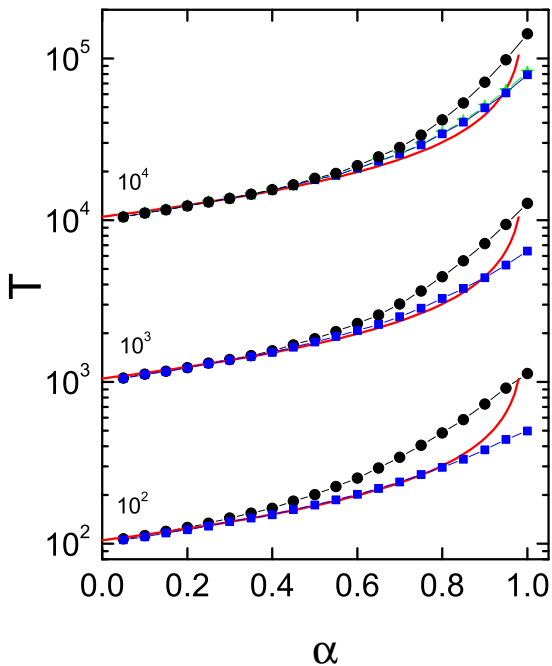}
\caption{MFPT as a function of $\alpha$ for $\nu=1$ (black points) and $\nu=100$ (blue squares). Green stars correspond to the case $\theta=-0.4$ and  
solid red lines mark the dependence (\ref{mfptf}). Barrier position $L$ is indicated for each bunch of curves.}
\end{figure}

\section{Position-dependent waiting time} 

The process described by Eq.(\ref{meq}) and (\ref{pv}) can be generalised to the case of the walker moving in a nonhomogeneous medium. 
The medium structure may influence, in particular, the waiting time distribution and we take into account this effect by making 
the rate $\nu$ position-dependent: $\nu=\nu(x)$. However, the representation of the master equation in terms of such a direct
generalisation of the fractional equation (\ref{lmeq1}) may not be valid. In particular, 
a strong decline of $\nu(x)$ so influences the relative importance of terms in the expansion of the master equation that it requires 
a qualitatively different approach and results in a different kind of the differential equation \cite{kam18}. 
That effect becomes clear when we assume $\nu(x)$ in a power-law form, 
\begin{equation}
\label{nuodx}
\nu(x)=\nu_0|x|^{-\theta}~~~~(\theta>-\alpha), 
\end{equation}
and the parameter $\theta$ serves as a measure of the medium structure nonhomogeneity. The constant $\nu_0$ was introduced for dimensional reasons; 
in the following, we set $\nu_0=1$. The form (\ref{nuodx}) of the waiting time spatial variability 
is natural, in particular, if the environment has a selfsimilar structure and was applied to describe fractals \cite{oshmet1}. 
If $\theta>\theta_{th}=1-\alpha$, we observe a dominance of the resting phase over the flying phase, in contrast to the case considered 
in Section III. Then 
the limits $s\to0$ and $k\to0$ must be taken simultaneously in such a way that $s/k$ remains constant which leads to a different 
mathematical description: the diffusion equation determines the density distribution, instead of the wave equation. 

Applying the above considerations to the walk in the bounded domain, one has to distinguish two forms of nonhomogeneity. 
The first form, that corresponds to $\theta<\theta_{th}$ and comprises 
both positive and negative $\theta$, we call 'weak nonhomogeneity'. This case is similar to the case discussed in the previous 
Section for the constant $\nu$ ($\theta=0$): the approximations leading to the Poisson equation (\ref{solv11a}) are valid 
and the flying phase prevails; consequently, MFPT is governed by Eq.(\ref{mfptf}). Fig.2 and 3 demonstrate that 
the numerically evaluated dependence of $T$ on both $L$ and $\alpha$ for a negative value of $\theta$ 
coincides with the results for $\nu=$ const. 

The 'strong nonhomogeneity' case ($\theta>\theta_{th}$) is characterised by a decreasing of the flying phase with time in the form 
of a power-law relaxation \cite{kam18} which means that after a long time-evolution particles predominantly stay in traps. 
If one introduces the absorbing barrier at a large distance from the initial point, the intensity of flying phase becomes very small 
before any particle reaches the barrier. Therefore, the escape process, MFPT in particular, is completely determined by $p_r(x,t)$. 
The process for large $\theta$ and distant barriers resembles the L\'evy flight since then the time spent by particle in traps 
strongly overbalances the time of flight: in the time scale imposed by the waiting time, 
the particle that leaves a trap almost instantaneously emerges at the barrier. 
\begin{figure}
\includegraphics[width=100mm]{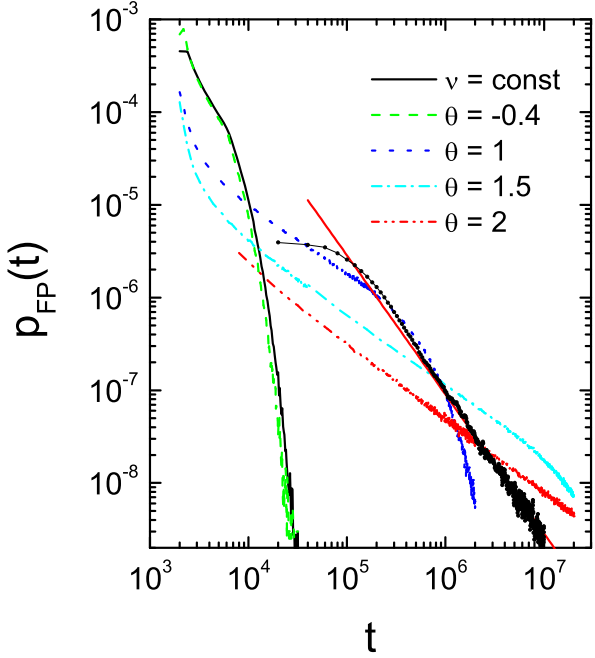}
\caption{First passage time distribution calculated from trajectory simulations for $\alpha=0.5$ and $L=2000$. 
Points correspond to the case of a single absorbing barrier at $L=10$ with $\nu=10^{-4}$ ($\alpha=0.5$ and $\theta=0$) 
and the curve assumes the shape $t^{-3/2}$ which is marked by a solid line.}
\end{figure}

Since for the fast diminishing $\nu(x)$ Eq.(\ref{lmeq1}) is no longer valid \cite{kam18}, our starting point is 
a generalisation of Eq.(\ref{solv10}), 
\begin{equation}
\begin{split} 
\label{solv10th}
p_v(k,s)=&c_1 \left[s^{\alpha-1}-s^{\alpha-3}\frac{1}{2}(1-\alpha)(2-\alpha)v^2k^2\right]\\
&\times\nu(x)p_r(k,s), 
\end{split} 
\end{equation}
where no specific form of $\nu(x)$ is assumed. A similar procedure as in the preceded Section yields the Poisson equation, 
\begin{equation}
\begin{split} 
\label{solv12th} 
\frac{\partial p_v(x,t)}{\partial t}=&c_1\left[\frac{\partial^2}{\partial t^2}+\frac{1}{2}(1-\alpha)(2-\alpha)v^2\frac{\partial^2}{\partial x^2}\right]\\
&\times{_0D}_{t}^{\alpha-2}\nu(x)p_r(x,t), 
\end{split} 
\end{equation}
that determines the quantity ${_0D}_{t}^{\alpha-2}\nu(x)p_r(x,t)$. Using the expansion (\ref{prwl}) and the already evaluated eigenfunctions, 
one can express the operator in Eq.(\ref{solv12th}) in the form, 
\begin{equation}
\label{poijsth}
\begin{split} 
&_0D_t^{\alpha-1}[\nu(x) X_n(x)T_n(t)]=\frac{C\pi(2n+1)}{2L}\\
&\times\left[a_n\exp(-\frac{C\pi(2n+1)t}{2L})+b_n\exp(\frac{C\pi(2n+1)t}{2L})\right]\\
&\times\cos\frac{\pi(2n+1)x}{2L}.
\end{split} 
  \end{equation} 
The coefficients $a_n$ and $b_n$ can be determined from conditions that we impose on $p_v(x,t)$: 
$p_v(x,0)=p_v(x,\infty)=0$; this yields $a_n=-2/C\pi(2n+1)$ and $b_n=0$. We derive the density $p_r(x,t)$ from Eq.(\ref{poijsth}) in two steps: 
first, we take the Laplace transform, 
   \begin{equation}
  \label{prlth}
\begin{split} 
s^{\alpha-1}\nu(x)p_r(x,s) =&\frac{1}{L}\sum_{n=0}^{\infty}\frac{1}{s+C\pi(2n+1)t/2L}\\
&\times\cos\frac{\pi(2n+1)x}{2L},
\end{split} 
  \end{equation}
and then, after multiplication by $s^{1-\alpha}$, invert the resulting expression. The final form of the density reads, 
  \begin{equation}
\begin{split} 
  \label{rfal1r}
p_r(x,t)=&\frac{1}{c_1\nu(x)L}\sum_{n=0}^{\infty}{_0}D_t^{1-\alpha}\exp(-\frac{C\pi(2n+1)t}{2L})\\
&\times\cos\frac{\pi (2n+1)x}{2L}, 
\end{split} 
  \end{equation}
where we used the relation $_0D_t^{-\alpha}\exp(0)=0$ \cite{milros}. 
\begin{figure}
\includegraphics[width=95mm]{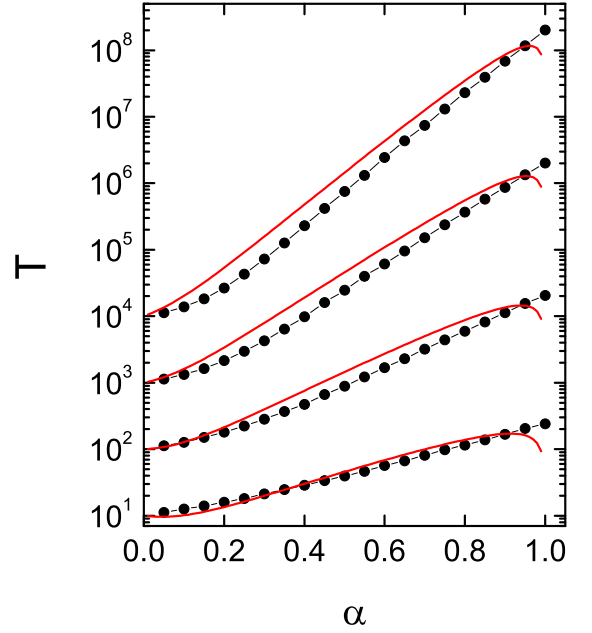}
\caption{MFPT as a function of $\alpha$ for $\theta=1$  and different values of $L$: 10, $10^2$, $10^3$ and $10^4$ (points, from bottom to top). Solid red lines mark the dependence (\ref{mfptfth1}).}
\end{figure}

To obtain the survival probability in a closed form, we have to assume a specific dependence $\nu(x)$; 
in the following, we assume Eq.(\ref{nuodx}). Then the integration of (\ref{rfal1r}) over $x$ yields the survival probability, 
 \begin{equation}
\begin{split} 
  \label{sodtfth}
S(t)=&\frac{2^{\theta+1}L^\theta}{\epsilon^\alpha\Gamma(1-\alpha)\pi^{\theta+1}}\sum_{n=0}^\infty {_0}D_t^{1-\alpha}
\exp(-\frac{C\pi(2n+1)t}{2L})\\
&\times\int_0^{(2n+1)\pi/2}u^\theta\cos u du, 
\end{split} 
  \end{equation}
where the leading term resolves itself to a Mellin-Ross function and then has the exponential asymptotics \cite{gormai,mathai2}. 
Fig.4 presents the numerically evaluated first passage time distribution $p_{FP}(t)$ (which is the derivative of $S(t)$) 
for a few values of $\theta$. We observe that if $\theta$ is large 
the exponential asymptotics is only present for very large values of $t$ while for smaller time a power-law segment emerges. 
The cases of constant $\nu$ and weak nonhomogeneity are also shown for comparison, the curves are similar 
and assume the exponential form (cf. Eq.(\ref{fptd})). To evaluate MFPT we have to take the integral from the Mellin-Ross 
function, ${_0}D_t^{1-\alpha}\exp(-\frac{C\pi(2n+1)t}{2L})$. After approximating this function by an exponential, the final result reads, 
  \begin{equation}
\begin{split} 
  \label{mfptfth}
T=&\frac{2^{\alpha+\theta+1}L^{\alpha+\theta}}{\epsilon^\alpha\Gamma(1-\alpha)C^\alpha \pi^{\alpha+\theta+1}} 
\sum_{n=0}^{\infty}\frac{1}{(2n+1)^{\alpha+\theta+1}}\\
&\times\int_0^{(2n+1)\pi/2}x^{\theta}\cos x dx. 
\end{split} 
  \end{equation} 
In order to obtain a simpler and more transparent expression for $T$ we estimate the integral from a mean value theorem 
(details are presented in Appendix) which procedure yields, 
\begin{equation}
\label{mfptfth1}
T=\frac{2^{\alpha+1}L^{\alpha+\theta}}{\epsilon^\alpha\Gamma(1-\alpha)(1+\theta)(C\pi)^\alpha}\sum_{n=0}^{\infty}\frac{(-1)^n}{(2n+1)^\alpha}.
  \end{equation}
A striking difference compared to the case of constant $\nu$ is the dependence $T(L)$: T rises faster than linear with $L$ 
and growth is stronger than for L\'evy flights. Fig.2 illustrates this result and compares Eq.(\ref{mfptfth1}) with 
the numerical calculations while the dependence $T(\alpha)$ is presented in Fig.5.

\section{Summary and conclusions}

We have discussed the L\'evy walk with random waiting times between displacements and derived time characteristics of the escape process from 
a domain bounded by two absorbing barriers. 
The combined density distribution for flights and rests, satisfying boundary conditions at barrier positions $\pm L$, 
has been evaluated by using solution of the Poisson equation; this equation 
determines a fractional operator from which the density evolution is derived. 
The simple expression for MFPT has been obtained and dependences on $L$ and $\alpha$ established. That result predicts, 
in particular, the proportionality of MFPT to $L$ which dependence is well-known from numerical analyses of the problem without rests. 
The mean waiting time $1/\nu$, that enters the model as a parameter, establishes the relative duration of resting and moving. 
Therefore, the model incorporates both the case of L\'evy walks process without rests (large $\nu$) and the limit $\nu\to0$ 
when the time of flight needed to reach the barrier becomes negligible compared to the resting time. This case reveals features 
typical for L\'evy flights, in particular, the dependence $MFPT\propto L^\alpha$. 
Another property of the L\'evy flights process, which can be observed when taking the limit $\nu\to0$ in L\'evy walks, is the validity of a Sparre-Andersen theorem. 
This theorem refers to escape from a domain which is open at one side and states that the first passage time distribution, $p_{FP}(t)$, 
should behave like $t^{-3/2}$ for any Markovian process \cite{met}. The numerical calculations reveal a power-law form of $p_{FP}(t)$ with slope rising with decreasing $\nu$; 
the form required by the Sparre-Andersen theorem is reached for $\nu=10^{-4}$ which is demonstrated at Fig.4. 
On the other hand, taking into account the finite 
waiting time also allowed us (since Eq.(\ref{mfptf}) does not depend on $\nu$) to analytically solve the first passage time problem 
for the case without rests which result had been unknown, to the best of our knowledge. 

The procedure has to be modified if one introduces a position dependence into the waiting time distribution 
which dependence is natural if the medium possesses a structure. For $\nu(x)$ falling sufficiently fast, the Poisson equation has been applied 
to derive the resting phase density since just this quantity determines the first passage time characteristics while the density of the 
flight phase dwindles with $L$. Then MFPT rises faster than linearly with $L$ and even faster than for the L\'evy flights process. 

\section*{APPENDIX} 

\setcounter{equation}{0}
\renewcommand{\theequation}{A\arabic{equation}} 

In the Appendix, we estimate the integral in Eq.(\ref{mfptfth}). The mean value theorem states that there exists such $c$ that, 
 $$
 \int_A f(x)g(x)dx=f(c)\int_A g(x)dx,
 $$
where $f(x)$ is a continuous function determined on a closed set $A$ and $g(x)$ is an integrable and nonnegative function. 
Let $f(x)=\cos x$ and $g(x)=|x|^{\theta}$. Then for any even $n$ we have $f(c)\in(0,1)$ while for any odd $n$ $f(c)\in(-1,0)$. 
Estimation of the integral for even $n$ by the upper bound and for odd $n$ by the lower bound yields 
$\int_0^{(2n+1)\pi/2}x^{\theta}\cos x dx\sim (-1)^n\int_0^{(2n+1)\pi/2}x^{\theta}dx$ and after evaluation of this integral 
we obtain the required estimation.


\begin{thebibliography}{99} 

\bibitem{gei}
T. Geisel, J. Nierwetberg, and A. Zacherl, Phys. Rev. Lett. {\bf 54}, 616 (1985). 

\bibitem{zum}
G. Zumofen and J. Klafter, Phys. Rev. E {\bf 47}, 851 (1993).

\bibitem{kla1}
J. Klafter, A. Blumen, and M. F. Shlesinger, Phys. Rev. A {\bf 35}, 3081 (1987).

\bibitem{zab}
V. Zaburdaev, S. Denisov, and J. Klafter, Rev. Mod. Phys. {\bf 87}, 483 (2015). 

\bibitem{froe}
D. Froemberg, M. Schmiedeberg, E. Barkai, and V. Zaburdaev, Phys. Rev E {\bf 91}, 022131 (2015). 

\bibitem{brok}
X. Brokmann, J.-P. Hermier, G. Messin, P. Desbiolles, J.-P. Bouchaud, and M. Dahan, 
Phys. Rev. Lett. {\bf 90}, 120601 (2003). 

\bibitem{marg}
G. Margolin and E. Barkai, Phys. Rev. Lett. {\bf 94}, 080601 (2005). 

\bibitem{kla2}
J. Klafter and G. Zumofen, Phys. Rev. E {\bf 49}, 4873 (1994).

\bibitem{zab1}
V. Yu. Zaburdaev and K. Chukbar, JETP {\bf 94}, 252 (2002). 

\bibitem{tay}
J. P. Taylor-King, E. van Loon, G. Rosser, and S. J. Chapman,
Bull. Math. Biol. {\bf 77}, 1213 (2015). 

\bibitem{kam17} 
A. Kami\'nska and T. Srokowski, Phys. Rev. E {\bf 96}, 032105 (2017). 

\bibitem{kam18} 
A. Kami\'nska and T. Srokowski, Phys. Rev. E {\bf 97}, 062120 (2018). 

\bibitem{red}
S. Redner, {\it A guide to first-passage processes} (Cambridge University Press, Cambridge, UK, 2001). 

\bibitem{dyb17} 
B. Dybiec, E. Gudowska-Nowak, E. Barkai, and A. A. Dubkov, Phys. Rev. E {\bf 95}, 052102 (2017). 


\bibitem{zoi}
A. Zoia, A. Rosso, and M. Kardar, Phys. Rev. E {\bf 76}, 021116 (2007). 

\bibitem{ryz}
I. S. Gradshteyn and I. M. Ryzhik, {\it Table of Integrals, Series, and Products} (Elsevier Inc., London, UK, 2007). 

\bibitem{schm}
M. Schmiedeberg, V. Y. Zaburdaev, and H. Stark, J. Stat. Mech. (2009) P12020. 

\bibitem{podl}
I. Podlubny {\it Fractional Differential Equations}
(Elsevier Academic Press, 1999). 

\bibitem{kilbas}
A. A. Kilbas, H. M. Srivastava, and J. J. Trujillo, 
{\it Theory and Applications of Fractional Differential Equations} (Elsevier, Amsterdam, 2006). 


\bibitem{oshmet1}
B. O'Shaughnessy and I. Procaccia, Phys. Rev. Lett. {\bf 54}, 455 (1985); 
R. Metzler, W. G. Gl\"ockle, and T. F. Nonnenmacher, Physica A {\bf 211}, 13 (1994).    

\bibitem{milros}
K. Miller, B. Ross, {\it An Introduction to the Fractional Calculus and Fractional Differential Equations}
(John Wiley \& Sons, Inc., 1993). 

\bibitem{gormai}
R. Gorenflo,  A. A. Kilbas,  F. Mainardi,  S. V. Rogosin
{\it Mittag-Leffler Functions, Related Topics}  (Springer-Verlag Berlin Heidelberg 2014).

\bibitem{mathai2}
A. M. Mathai,  H. J. Haubold, {\it Special Functions for Applied Scientists}
(Springer, New York, 2008).

\bibitem{met} 
R. Metzler and J. Klafter, J. Phys. A: Math. Gen. {\bf 37}, R161 (2004). 

\end{thebibliography}
\end{document}